\documentclass[a4paper]{jpconf}
\usepackage{graphicx}

\usepackage[breaklinks]{hyperref}
\usepackage{booktabs}
\usepackage{graphicx,amsmath,amssymb}

\begin{document}

\title{Mind the Resonances: Final stages of accretion into bumpy black holes}

\author{Georgios Lukes-Gerakopoulos$^1$ and George Contopoulos$^2$}

\address{$^1$ Theoretical Physics Institute, University of Jena,
\\ 07743 Jena, Germany}
\address{$^2$  Research Center for Astronomy and Applied Mathematics, Academy of
Athens,\\ Soranou Efesiou 4, GR-11527 Athens, Greece}

\ead{gglukes@gmail.com}

\begin{abstract}
 In this article we discuss a possible way of testing the Kerr black hole
 hypothesis by taking advantage of phenomena correlated with chaotic motion in the 
 final stages of an accretion disk around a bumpy black hole. We anticipate that
 these phenomena should have an imprint in the electromagnetic spectrum coming from
 the accretion disk. 
\end{abstract}

\section{Introduction}

The gravitational wave detectors are anticipated to provide observations which will
help us to study massive compact objects and to test our theoretical expectations
about their nature. Namely, the spacetime around black holes is expected to be
described by the Kerr metric. One way to test the Kerr hypothesis is to search in
the gravitational wave signals for any deviation from the corresponding signals we
would expect to come from a Kerr spacetime background, see e.g.,
\cite{eLISA,Bambi11,Johan12a} and references therein. However, alternative tests
for our expectation might be based on contemporary plausible astronomical
observations. Namely, the Kerr hypothesis can be tested by studying the
electromagnetic spectra coming from a black hole's surrounding environment (e.g., 
\cite{Bambi12a,Bambi12b,Johan10a,Johan10b,Johan11a,Johan11b,Johan12b,Pappas12}). 

Both gravitational and electromagnetic Kerr hypothesis tests are usually applied on
axially symmetric perturbations of the Kerr spacetime, such spacetimes are often
called in the literature bumpy black holes (name given in \cite{Collins04}) or
non-Kerr spacetimes. By perturbing the Kerr spacetime it seems that in general we
destroy a generalized Noether symmetry (e.g., \cite{Markakis12} and references
therein) responsible for the existence of the Carter constant \cite{Carter68}. Even
if there are attempts of finding higher order Killing tensors expressing this
symmetry in specific bumpy black hole spacetimes (\cite{Brink11} and references
therein), the numerical results show that chaos appears when we examine geodesic
orbits in these spacetimes \cite{Gair08,LGAC10,CLGA11,CHL12,LG12}, which in turn
implies that these spacetimes correspond to a non-integrable system and therefore
no Carter-like constant can exist. On the other hand, if we demand the preservation
of the Carter constant after the perturbation, it seems that we have to relax the
requirement that the perturbed metric satisfies the Einstein equations
\cite{Vigeland11}. Thus, it appears that after the perturbation of the Kerr metric
we have to either lose the Carter constant or drop the requirement that the new
metric has to satisfy the Einstein equations.  

The non-integrability of the bumpy black holes is correlated with several
interesting aspects of classical non-integrable systems. In particular, in
\cite{CHL12} we have studied how the bifurcation of the periodic orbits at the 
resonances lead to escaping orbits in a classical system of two coupled harmonic
oscillators and a relativistic system described by a subclass of the Manko-Novikov
(MN) metric family \cite{ManNov92}. In the present article we attempt to give a
physical interpretation to the study done in \cite{CHL12}. Namely, we try to
interpret our findings as possible imprints in the electromagnetic spectra
indicating a method to distinguish bumpy black hole spacetimes from Kerr spacetimes
by exploiting the non-linear phenomena studied in \cite{CHL12} at the final stages
of accretion into a non-Kerr compact object, in a similar way as done by the
authors of \cite{BambiBar11}.

The article is organized as follows. We introduce the MN spacetime in section
\ref{sec:MNspa}. Section \ref{sec:MNgeo} summarizes basic elements of geodesic
motion in axisymmetric and stationary spacetimes. We discuss, by interpreting the 
numerical examples given in \cite{CHL12}, the final stages of accreting matter into
a MN bumpy black hole in section \ref{sec:AccDisk}. Section \ref{sec:concl} briefs
our main conclusion.
  
\section{The Manko-Novikov spacetime} \label{sec:MNspa}   

The bumpy black hole spacetime we used in \cite{CHL12} is a spacetime which
belongs to the so-called Manko-Novikov (MN) metric family \cite{ManNov92}.
Manko and Novikov found an exact vacuum solution of Einstein's equations which
describes a stationary, axisymmetric, and asymptotically flat spacetime with
arbitrary mass-multipole moments \cite{ManNov92}. The MN metric subclass we used
was introduced by \cite{Gair08} and deviates from the Kerr by the anomalous
quadrupole moment $q$. The quantity $q$ measures how much the MN quadrupole moment
$Q$ departs from the Kerr quadrupole moment $Q_{Kerr}=-S^2/M$ (i.e.,
$q=(Q_{Kerr}-Q)/M^3$), where $M$ and $S$ are the mass and spin of a Kerr black
hole. The line element of the MN metric in the Weyl-Papapetrou cylindrical
coordinates $(\rho, \varphi, z)$ is 
 \begin{equation} \label{eq:MNmetric}
  ds^2=-f(dt-\omega d\varphi)^2 + f^{-1} [e^{2\gamma} (d\rho^2 + dz^2)
       +\rho^2 d\varphi^2],
 \end{equation}
where $f,~\omega,~\gamma$ are considered as functions of the prolate spheroidal
coordinates $v, w$, while the coordinates $\rho,z$ can be expressed as functions
of $v, w$ as well. Thus
\begin{equation}\label{func:trans}
 \rho=k \sqrt{(v^2-1)(1-w^2)},\quad z=k v w,
\end{equation}
and
\begin{subequations}
\begin{eqnarray}
 f &=& e^{2 \psi}\frac{A}{B}, \label{ffunc} \\
 \omega &=& 2 k e^{-2 \psi}\frac{C}{A}-4 k \frac{\alpha}{1-\alpha^2}, \\
 e^{2 \gamma} &=& e^{2 \gamma^\prime}\frac{A}{(v^2-1)(1-\alpha^2)^2},
 \label{fexpgam} \\
 A &=& (v^2-1)(1+a~b)^2-(1-w^2)(b-a)^2,\label{fA} \\
 B &=& [(v+1)+(v-1)a~b]^2+[(1+w)a+(1-w)b]^2,\label{fB} \\
 C &=& (v^2-1)(1+a~b)[(b-a)-w(a+b)] \nonumber \\
   &&+ (1-w^2)(b-a)[(1+a~b)+v(1-a~b)], \\
 \psi &=& \beta \frac{P_2}{R^3}, \label{fC}\\
 \gamma^\prime &=& \ln{\sqrt{\frac{v^2-1}{v^2-w^2}}}+\frac{3\beta^2}{2 R^6}
 (P_3^2-P_2^2) \nonumber \\ &+& \beta \left(-2+\displaystyle{\sum_{\ell=0}^2}
 \frac{v-w+(-1)^{2-\ell}(v+w)}{R^{\ell+1}}P_\ell\right), \label{fgampr}%\\
\end{eqnarray}
\end{subequations}

\begin{subequations}
\begin{eqnarray}
 a &=& -\alpha \exp {\left[-2\beta\left(-1+\displaystyle{\sum_{\ell=0}^2}
 \frac{(v-w)P_\ell}{R^{\ell+1}}\right)\right]}, \label{fa}\\
 b &=& \alpha \exp {\left[2\beta\left(1+\displaystyle{\sum_{\ell=0}^2}
 \frac{(-1)^{3-\ell}(v+w)P_\ell}{R^{\ell+1}}\right)\right]}, \label{fb}\\
 R      &=& \sqrt{v^2+w^2-1}, \label{fR}\\
 P_\ell &=& P_\ell (\frac{v~w}{R}). \label{fLegA}
\end{eqnarray}
\end{subequations}
Here $P_\ell(\zeta)$ is the Legendre polynomial of order $l$
\begin{equation} \label{fLeg}
 P_\ell(\zeta)=\frac{1}{2^\ell \ell!}
\left(\frac{d}{d\zeta}\right)^\ell(\zeta^2-1)^\ell,
\end{equation}
while the parameters $k,\alpha,\beta$ are related to the mass $M$, the spin $S$,
and the quadrupole deviation $q$ through the expressions
\begin{equation}
\begin{array}{r}
\alpha=\frac{-1+\sqrt{1-\chi^2}}{\chi},
\end{array}
\begin{array}{c}
k=M\frac{1-\alpha^2}{1+\alpha^2},
\end{array}
\begin{array}{l}
\beta=q \left( \frac{1+\alpha^2}{1-\alpha^2} \right)^3.
\end{array}
\label{freepar}
\end{equation}
while $\chi$ is the dimensionless spin parameter $\chi=S/M^2$. These formulae
give the Kerr metric for $q=0$. We use the geometrical units $c=G=1$ throughout the
article, and without loss of generality we have set also $M=1$.

\section{Geodesic Motion in the Manko-Novikov spacetime} \label{sec:MNgeo}  

The geodesic orbits of a test particle of mass $\mu$ are described as equations
of motion of the following Lagrangian
\begin{equation}
L=\frac{1}{2}~\mu~g_{\mu\nu}~ \dot{x}^{\mu} \dot{x}^{\nu},
\label{LagDef}
\end{equation}
where the dots mean derivatives with respect to the proper time. Except from the
Lagrangian itself $L=-\mu$, the MN metric has two more integrals of motion, namely
the energy (per unit mass)
\begin{equation} \label{eq:EnCon}
E=-\frac{\partial L}{\partial \dot{t}}/\mu=
f (\dot{t} - \omega~ \dot{\varphi}),
\end{equation}
and the z-component of the angular momentum (per unit mass)
\begin{equation} \label{eq:AnMomCon}
L_z =\frac{\partial L}{\partial \dot{\varphi}}/\mu=
f \omega (\dot{t} - \omega~ \dot{\varphi})+ f^{-1} \rho^2 \dot{\varphi},
\end{equation}
The Kerr metric has one more integral of motion which is independent and in
involution with the other integrals, the so-called Carter constant \cite{Carter68}.
Thus the Kerr metric corresponds to an integrable system. However, the MN model
misses such a constant, which means that MN corresponds to a non-integrable system,
and therefore chaos may appear.

By simple algebraic manipulations of the metric (\ref{eq:MNmetric}) and the eqs.
(\ref{eq:EnCon}), (\ref{eq:AnMomCon}) we find that the MN system satisfies the
relation
\begin{equation} \label{eq:Veff}
\frac{1}{2} (\dot{\rho}^2 + \dot{z}^{2}) + V_\mathrm{eff} (\rho, z)=0,
\end{equation}
where the effective potential $V_\mathrm{eff} (\rho, z)$ depends on $q$, $E$ and
$L_z$, and the geodesic motion can be restricted on a meridian plane
$(\varphi = \mathrm{const})$. Moreover, the motion takes place inside a region
defined by $V_\mathrm{eff} (\rho, z)\le 0$, the border of which is the curve of
zero velocity (CZV) 
\begin{equation} \label{eq:CZV}
 V_\mathrm{eff} \equiv \frac{1}{2} e^{-2\gamma}\left[f-E^2+\left(\frac{f}{\rho}
 (L_z-\omega E) \right)^2\right] = 0.
\end{equation}

\begin{figure}[htp]
\begin{minipage}{\textwidth}
\includegraphics[width=0.425\textwidth]{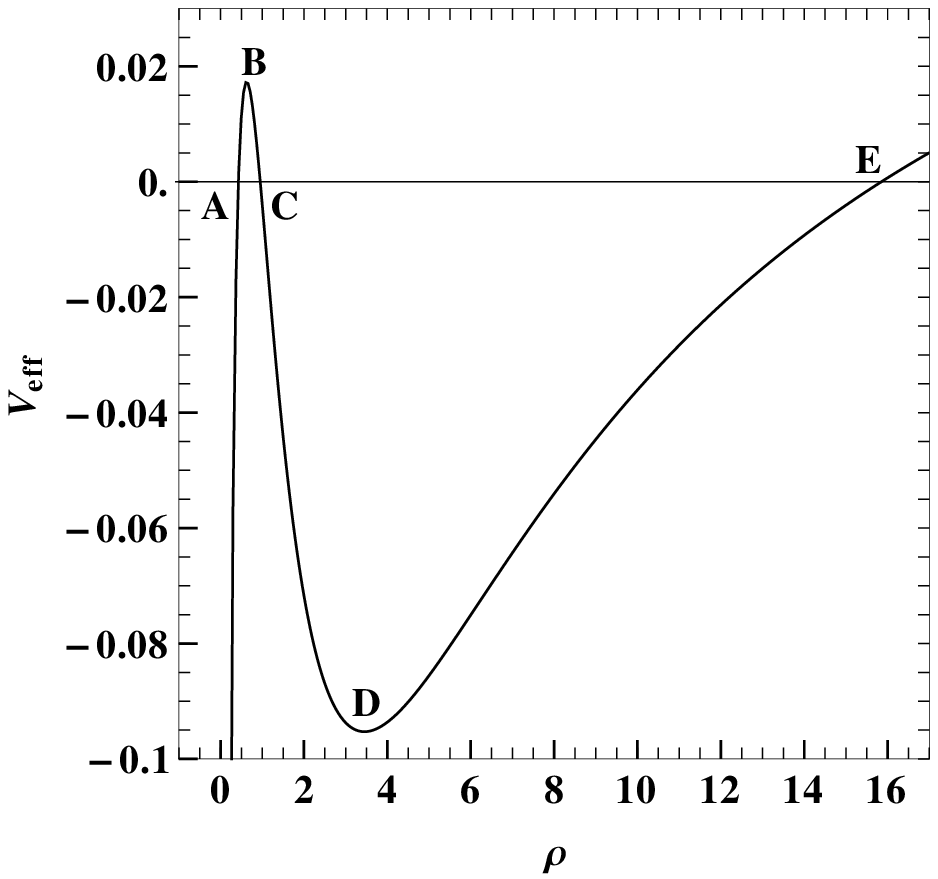}
\includegraphics[width=0.400\textwidth]{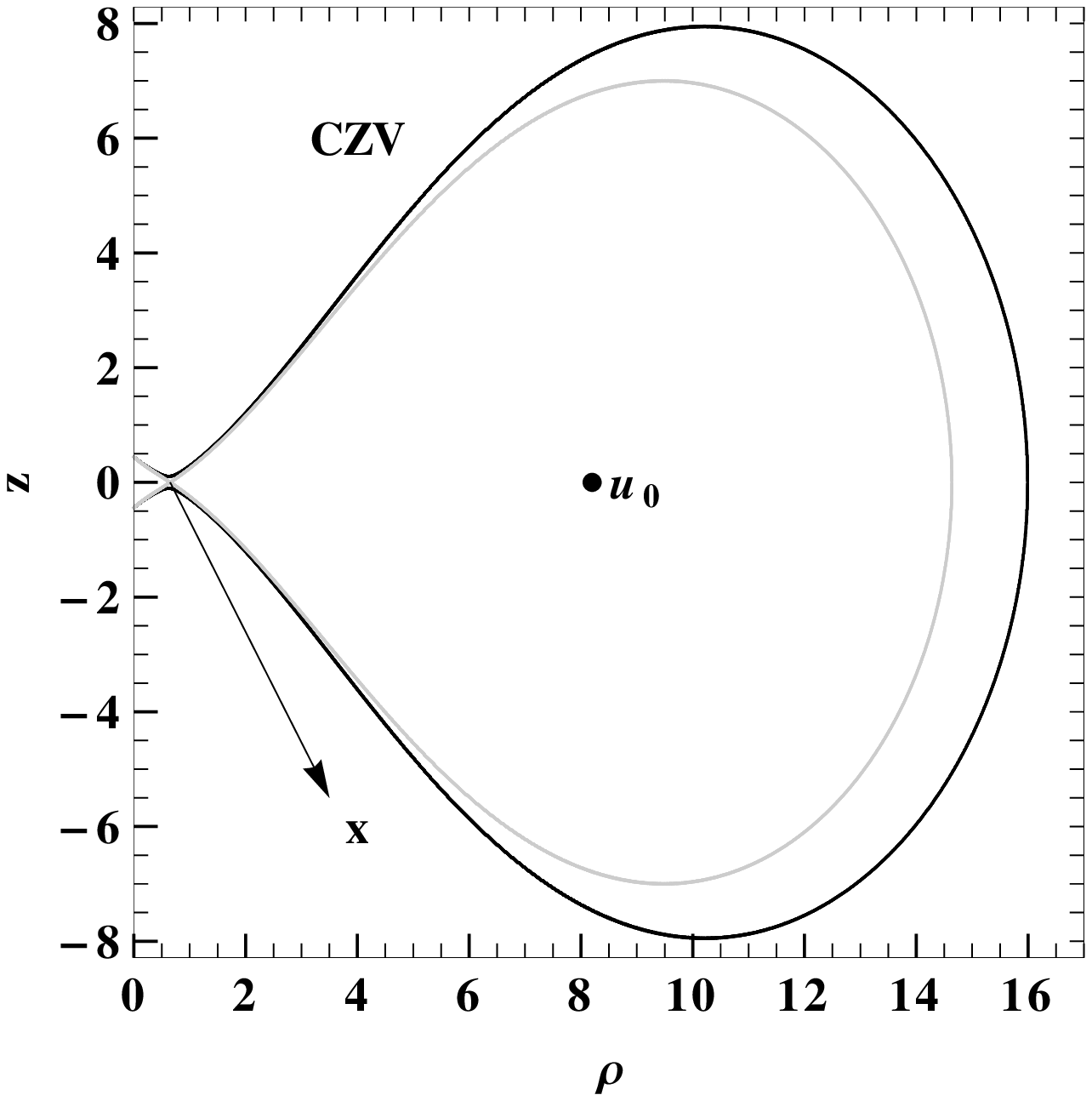}
\caption{\label{FigKerrCZV} The left panel shows the portrait of the effective
potential $V_\mathrm{eff}$ along the equatorial line ($z=0$) for $E=0.95$,
$L_z=2.5$, $\chi=0.9$, and $q=0$ (Kerr metric). The horizontal line
$V_\mathrm{eff}=0$ indicates the upper limit reached by the non-escaping to
infinity orbits, while the letters A, C, and E indicate the position of the roots
of $V_\mathrm{eff}$. Letters B, and D indicate the position of local extrema of
$V_\mathrm{eff}$. The right panel shows the CZV (black curve) on the meridian plane
$(\rho,~z)$ for a slightly lower angular momentum than in the left panel
($L_z=2.46$), the separatrix (gray curve) emanating from the unstable periodic
orbit labeled by the letter ``x'' and the crossing of a stable periodic orbit
through the $z=0$ line (big dot) labeled by $\mathbf{u}_0$.}
\end{minipage}
\end{figure}

From the shape of the effective potential (\ref{eq:CZV}) we can discern the basic
dynamics of the system like in a classical system. For instance in the integrable
case of the Kerr metric the roots of the effective potential provide the boundaries
of the non-escaping to infinity orbits (A, C, E in the left panel of Fig.
\ref{FigKerrCZV}). Between C and E lie the bounded non-plunging orbits, while from
A leftwards lie the orbits that are plunging to the black hole (left panel of Fig.
\ref{FigKerrCZV}). Moreover, the local extrema of $V_\mathrm{eff}$ implies the
existence of simple periodic orbits, for example the local minimum D indicates the
existence of a simple stable periodic orbit indicated by $\mathbf{u}_0$ in the
right panel of Fig. \ref{FigKerrCZV}, while the local maximum B implies that a
simple unstable periodic orbit will appear (x in the right panel of Fig.
\ref{FigKerrCZV}) for a specific range of $V_\mathrm{eff}$ parameters, namely if we
lower the $L_z$ in the case shown in the left panel of \ref{FigKerrCZV} to
$L_z=2.46$. When the local maximum B becomes negative, the separation between
plunging and non-plunging orbits is defined by the separatrix (gray curve in the
right panel) emanating from the unstable periodic orbit x. The orbits which lay
between $\mathbf{u}_0$ and the part of the separatrix encircling $\mathbf{u}_0$ are
bounded non-plunging orbits, all the other orbits inside the CZV plunge into the
black hole. For a specific set of $E$ and $L_z$ values (lower than those in Fig.
\ref{FigKerrCZV}), when all the five points indicated by the letters shown in the
left panel of Fig. \ref{FigKerrCZV} have merged into one point, the periodic orbits
x and $\mathbf{u}_0$ merge into an indifferently stable periodic orbit, which is
known in the bibliography as the Innermost Stable Circular Orbit (ISCO). Inside
from an ISCO (leftwards in Fig. \ref{FigKerrCZV}) only plunging orbits exist.

\begin{figure}[htp]
\begin{minipage}{\textwidth}
\includegraphics[width=0.425\textwidth]{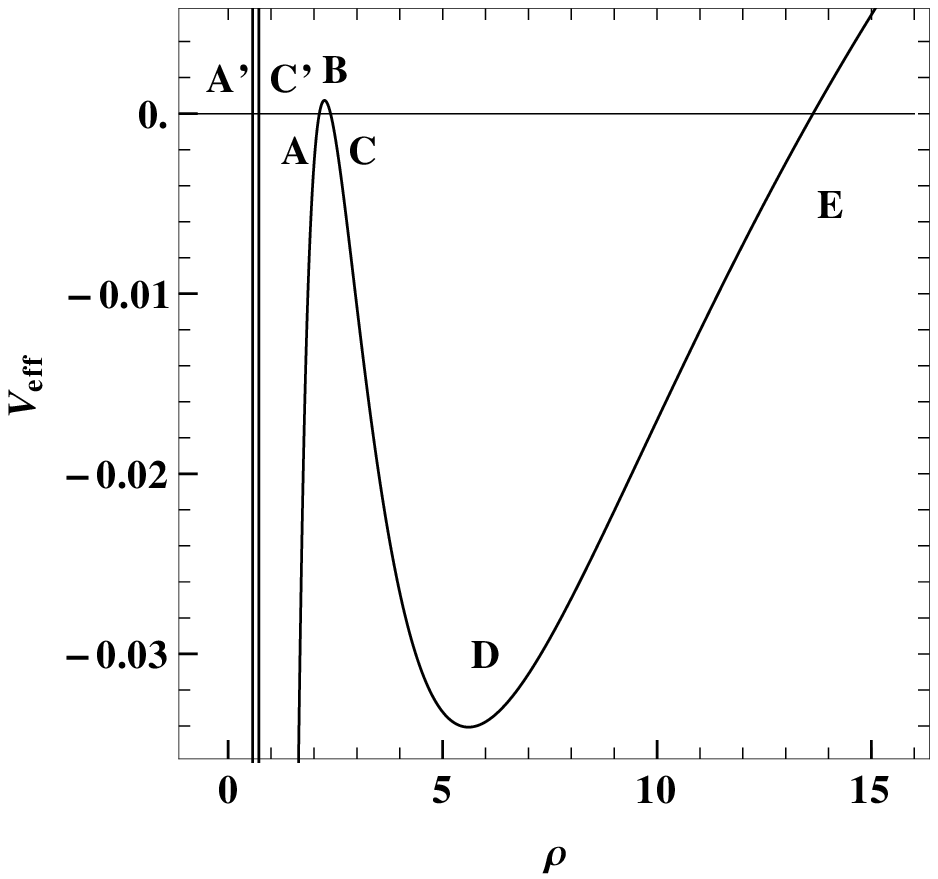}
\includegraphics[width=0.400\textwidth]{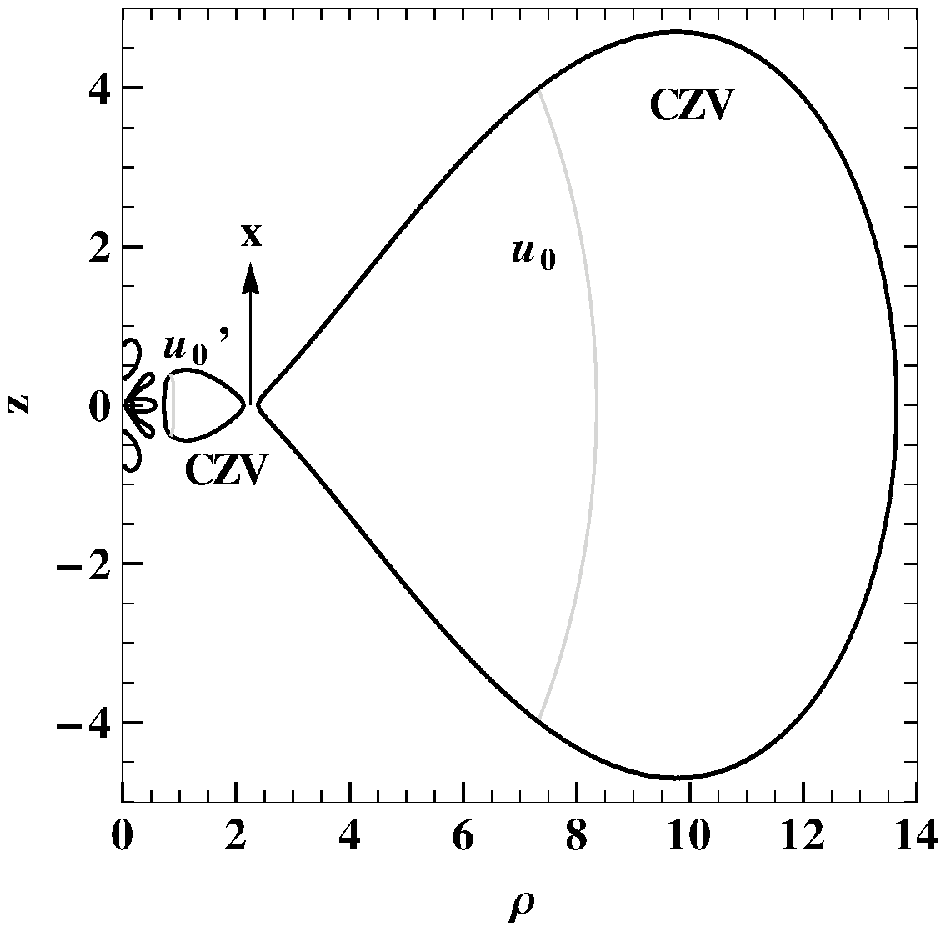}
\caption{\label{FigMNCZV} The left panel shows the portrait of the effective
potential $V_\mathrm{eff}$ along the equatorial line ($z=0$) for $E=0.95$, $L_z=3$,
$\chi=0.9$, and $q=0.95$. The letters $\mathrm{A}^\prime$, $\mathrm{C}^\prime$, A,
C and E indicate the position of the roots of $V_\mathrm{eff}$. Letters B, and D
indicate the position of local extrema of $V_\mathrm{eff}$. The right panel shows
the CZVs (black curves) on the meridian plane $(\rho,~z)$, and the projections on
the plane of the simple stable periodic orbits $\mathbf{u}_0$, $\mathbf{u}_0^\prime$
(gray curves). While $x$ indicates the region where an unstable periodic orbit will
appear for lower angular momentum $L_z$, or greater energy $E$.}
\end{minipage}
\end{figure}

In the case of the subclass of the MN family and for the parameters we have 
examined, the $V_\mathrm{eff}$ takes a little bit different shape (left panel of
Fig. \ref{FigMNCZV}). Except from the roots we have seen in the Kerr case, two
extra root appear the $\mathrm{A}^\prime$ and $\mathrm{C}^\prime$. Hence, an extra
non-plunging region between $\mathrm{C}^\prime$ and A appear. The local minimum
between $\mathrm{C}^\prime$ and A is extremely low (far out of the range shown in
Fig. \ref{FigMNCZV}) and it indicates the existence of a new simple stable periodic
orbit ($\mathbf{u}_0^\prime$ in the right panel of Fig. \ref{FigMNCZV}). B and D
again indicate a simple unstable periodic orbit (when B is below $V_\mathrm{eff}$)
and a simple stable periodic orbit (x and $\mathbf{u}_0$ in the right panel of Fig.
\ref{FigMNCZV}) respectively. Similarly as in the Kerr case, for a specific set of
$E$ and $L_z$ values, when all the five points indicated by the not primed letters
shown in the left panel of Fig.  \ref{FigMNCZV} have merged into one point, the
periodic orbits x and $\mathbf{u}_0$ merge into an indifferently stable periodic
orbit, which although it is not the innermost stable circular orbit, we are going
to call it ISCO, like the authors of \cite{BambiBar11} did. Moreover, five
leaf-like plunging regions appear on the left side of the right panel of Fig. 
\ref{FigMNCZV}.  

\begin{figure}[htp]
\includegraphics[width=0.4\textwidth]{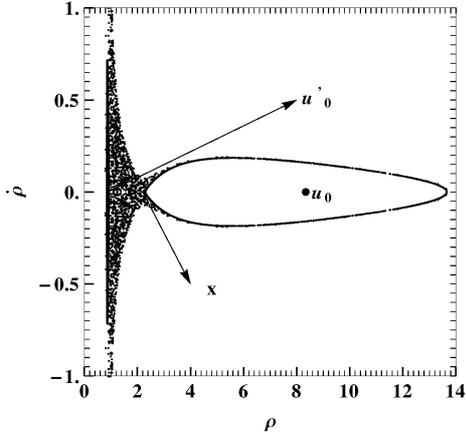}
\hspace{0.1\textwidth}
\begin{minipage}[b]{0.4\textwidth}
\caption{\label{FigSSLz2d98} A Poincar\'{e} section $(\rho, \dot{\rho})$ along the
$z=0$ for $E=0.95$, $L_z=2.98$, $\chi=0.9$, and $q=0.95$. The position of the
simple periodic orbit $\mathbf{u}_0$ is shown by a big dot, while the positions of
the simple unstable periodic orbit x and of the simple periodic orbit
$\mathbf{u}_0^\prime$ are shown by arrows. }
\end{minipage}
\end{figure}

If we cut the phase space along the along the equatorial plane ($z=0$) plane and
keep only the orbits crossing it with positive velocity, we produce a surface of
section, also known as a Poincar\'{e} section. In Fig. \ref{FigSSLz2d98} we show
such a section and the periodic orbits x, $\mathbf{u}_0^\prime$ and $\mathbf{u}_0$
on it. The scattered points seen in Fig. \ref{FigSSLz2d98} belong to a chaotic
region. Because the MN corresponds to a non-integrable system, this chaotic region
is swept densely by the asymptotic manifolds emanating from x instead of the
separatrix of the integrable case.   

\section{The final stages of an accretion disk in a Manko-Novikov spacetime} 
\label{sec:AccDisk}  

Bambi and Barausse in \cite{BambiBar11} found that during the accretion process in
MN spacetime the gas crossing the ISCO can follow four different types of plunge: 
\begin{enumerate}
 \item The gas plunges radially into the compact object like in the Kerr case. 
 \item The gas plunges radially, but gets trapped in a region near the compact
  object.
 \item The gas plunges to the central object due to a vertical to the equator
 instability.
 \item The gas plunges due to a vertical to the equator instability, but it gets
 trapped to two symmetrical to the equator regions.  
\end{enumerate}

Our study in a way follows up the second scenario, and explores what happens with
the entrapped gas. The gas in our simple model consists of collisionless ``test
particle'' fluid. We neglect the self-gravity of the gas, and the particles track
the geodesic orbits of the MN spacetime background. The above approximation might
seem a little irrelevant for a gas where the free path is short. However, when the
gas particles collide, they just ``exchange'' their position in the phase space; in
a great view they all still trace the background on which they are moving even if
each of them individually does not follow geodesic orbit, but rather a great
collection of small geodesic paths. In fact, we are not so interested on each
particle's individual orbit, but rather we are concerned with the escape rate of a
``swarm'' of particles, thus, in our simple approximation we allow ourselves to
ignore the collisions. Moreover, in our scenario the particles of the accretion
matter share a constant angular momentum $L_z=3$, but they radiate energy $E$ away.
This setup is like having a disk which radiates without losing its angular momentum.

\begin{figure}[htp]
\begin{minipage}{\textwidth}
\includegraphics[width=0.33\textwidth]{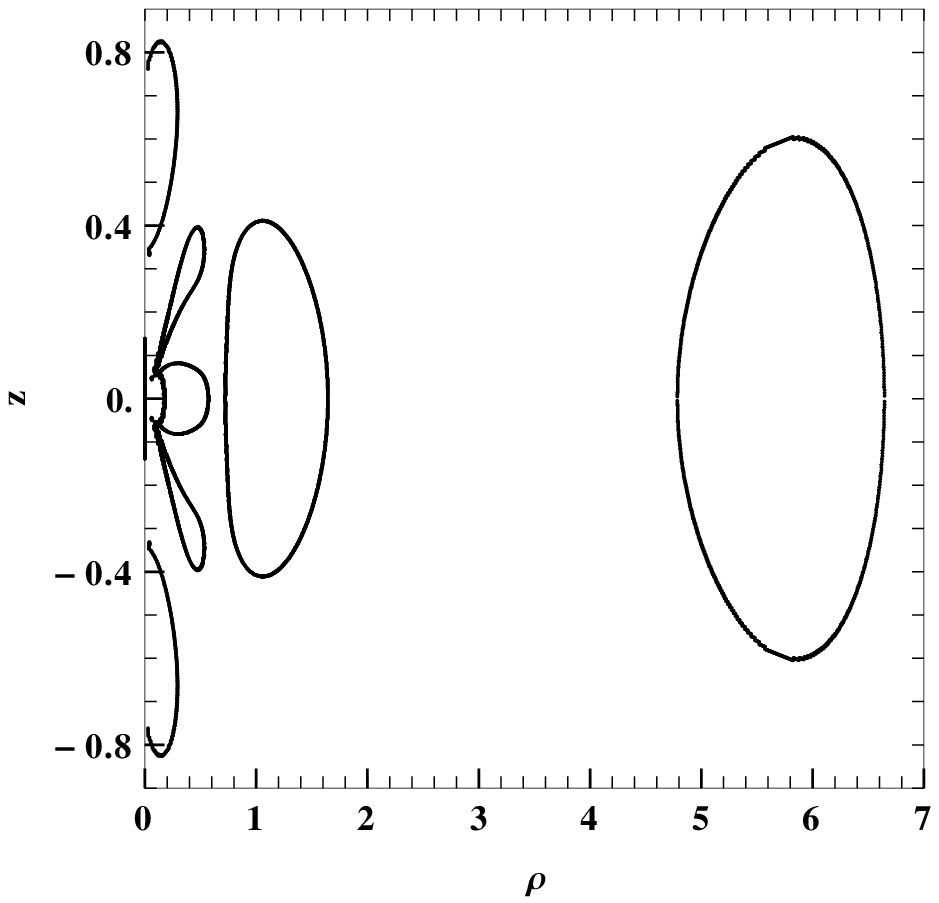}
\includegraphics[width=0.33\textwidth]{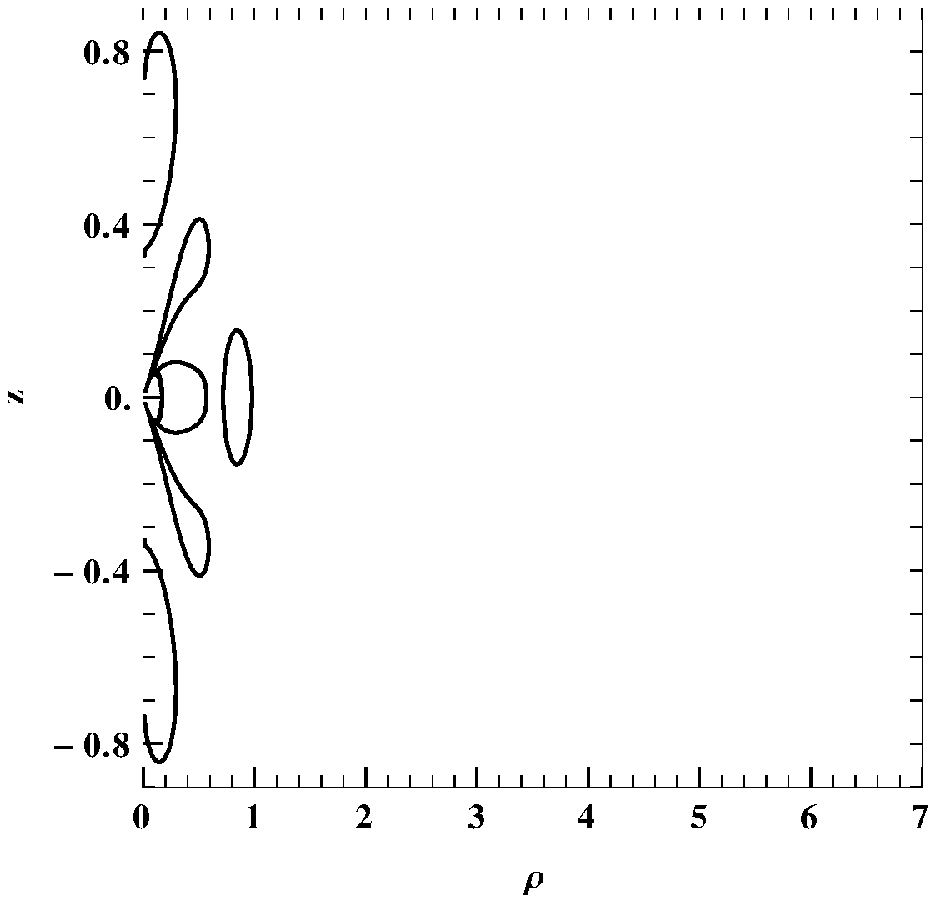}
\includegraphics[width=0.33\textwidth]{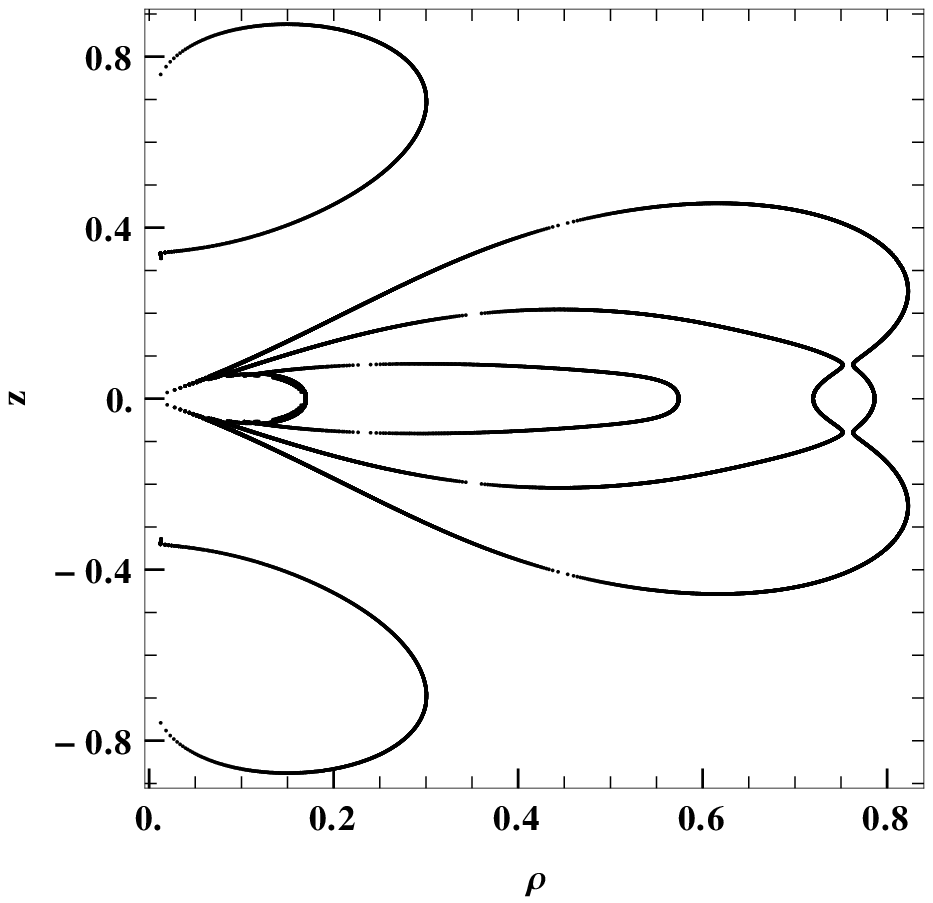}
\caption{\label{FigRe} The CZVs as we reduce energy from the left panel to the right
panel for constant $L_z =3$ (beginning from left $E=0.933$, $E=0.7$, $E=0.1899$).}
\end{minipage}
\end{figure}

We investigated the MN spacetime background for the values $q=0.95$, and $\chi=0.9$. 
By keeping the z-angular momentum $L_z =3$ fixed and reducing the energy, the outer
region around $\mathbf{u}_0$ (right panel of Fig. \ref{FigMNCZV}) gradually shrinks
(left panel of Fig. \ref{FigRe}), until it disappears and only the inner region
around  $\mathbf{u}_0^\prime$ remains (central panel of Fig. \ref{FigRe}). During
the energy reduction the leaf like plunging region on either side of the inner
region expand rightwards towards the inner region, until they reach it (right panel
of Fig. \ref{FigRe}). Our study in \cite{CHL12} focuses on the phase space
transition from having bounded non-plunging orbits to getting plunging orbits. One
must keep in mind that all the different energy states coexist, and the particles
slowly drift with different rates of energy loss towards the inner regions.

\begin{figure}[htp]
\begin{minipage}{\textwidth}
\includegraphics[width=\textwidth]{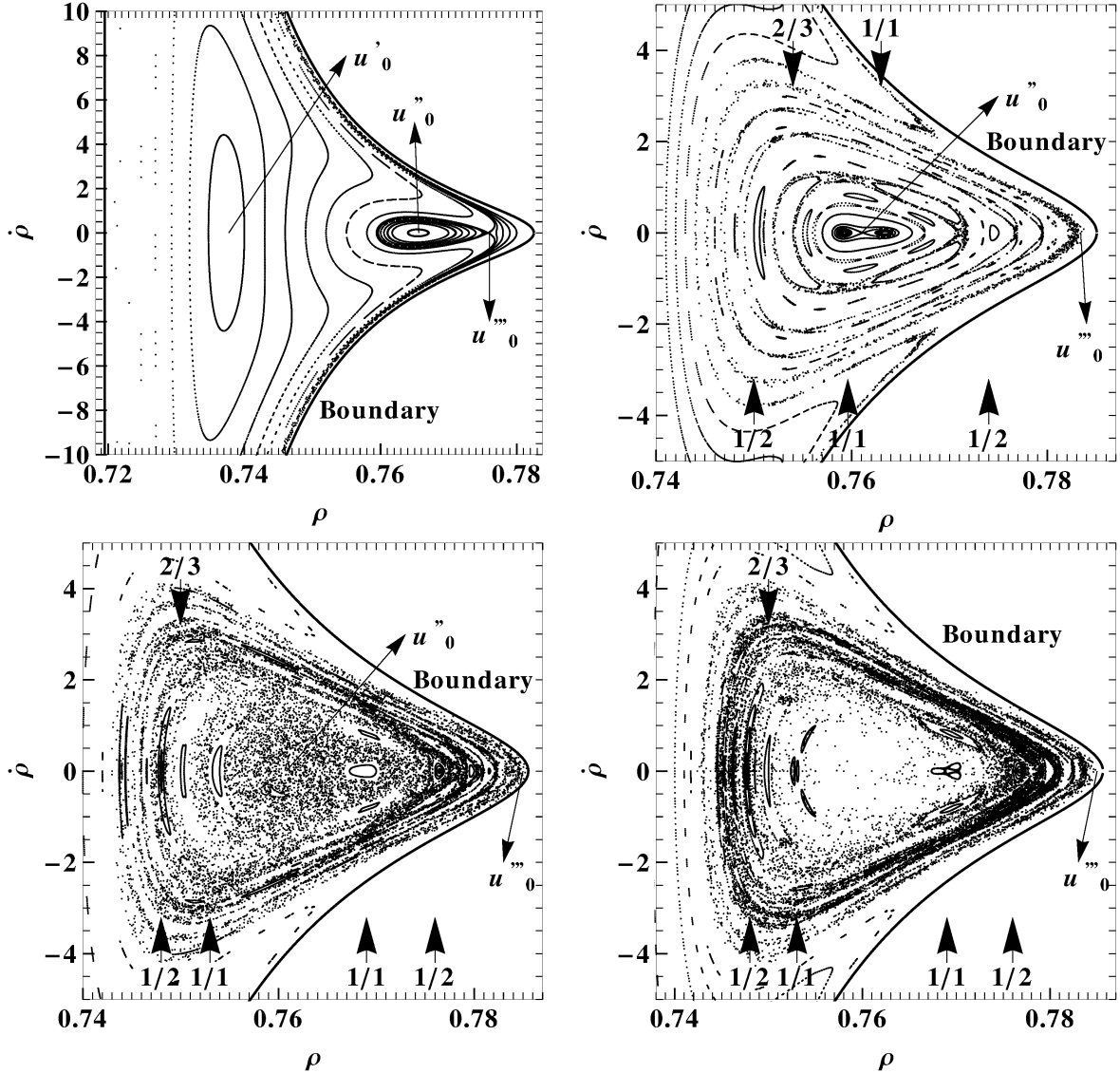}
\caption{\label{FigSS} Surfaces of section $(\rho,\dot{\rho})$ for different
energies: upper left panel $E=0.192$, upper right panel $E=0.19045$, lower left 
panel $E=0.19015$ and lower right panel $E =0.1901$. On the plots the simple
periodic orbits $\mathbf{u}_0^\prime$, $\mathbf{u}_0^{\prime\prime}$,
$\mathbf{u}_0^{\prime\prime\prime}$ and the positions on the $\dot{\rho}=0$ line of
low order periodic orbits $1/1$, $1/2$, $2/3$ are labeled.  }
\end{minipage}
\end{figure}

It is quite interesting to monitor the phase space transition of the inner region
from a region containing non-plunging orbits to a region with plunging orbits.
Initially the region is occupied mostly by regular orbits, but about $E=0.192$ a
tangent bifurcation produces a couple of new simple periodic orbits
$\mathbf{u}_0^{\prime\prime}$, $\mathbf{u}_0^{\prime\prime\prime}$ (upper left
panel of Fig. \ref{FigSS}), a stable and an unstable one respectively. The periodic
orbit $\mathbf{u}_0^{\prime\prime}$ changes its stability to instability and vice
versa an infinite number of times, while periodic orbits, like the low order ones
seen in the upper right panel of Fig. \ref{FigSS}, further bifurcate from it. This
leads to a gradually increasing volume occupied by chaotic orbits (lower left panel
of Fig. \ref{FigSS}), and when the CZV around the inner region is reached by the
leaf-like plunging region (right panel of Fig. \ref{FigRe}), then the chaotic
orbits lying around $\mathbf{u}_0^{\prime\prime}$ become plunging (lower right
panel of Fig. \ref{FigSS}). As the energy drops the period of
$\mathbf{u}_0^{\prime\prime}$ increases and tends to infinity. In fact it becomes
actually infinite when the CZV of the inner region opens up (see \cite{CHL12}). In
a way the orbit of $\mathbf{u}_0^{\prime\prime}$ ``traces'' the path to the chaotic
orbits in order for them to plunge.

\begin{figure}[htp]
\begin{minipage}{\textwidth}
\includegraphics[width=0.5\textwidth]{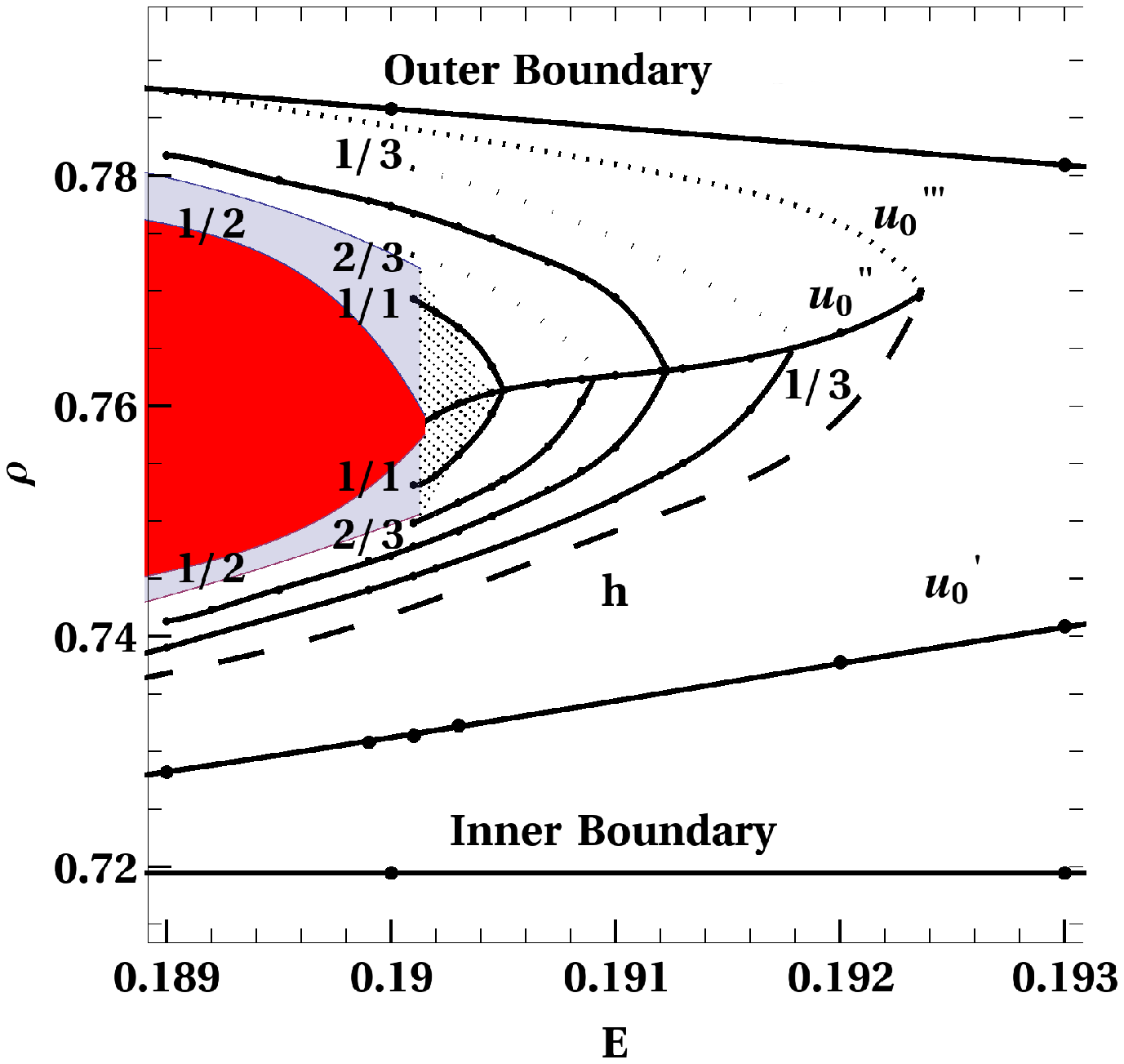}
\includegraphics[width=0.5\textwidth]{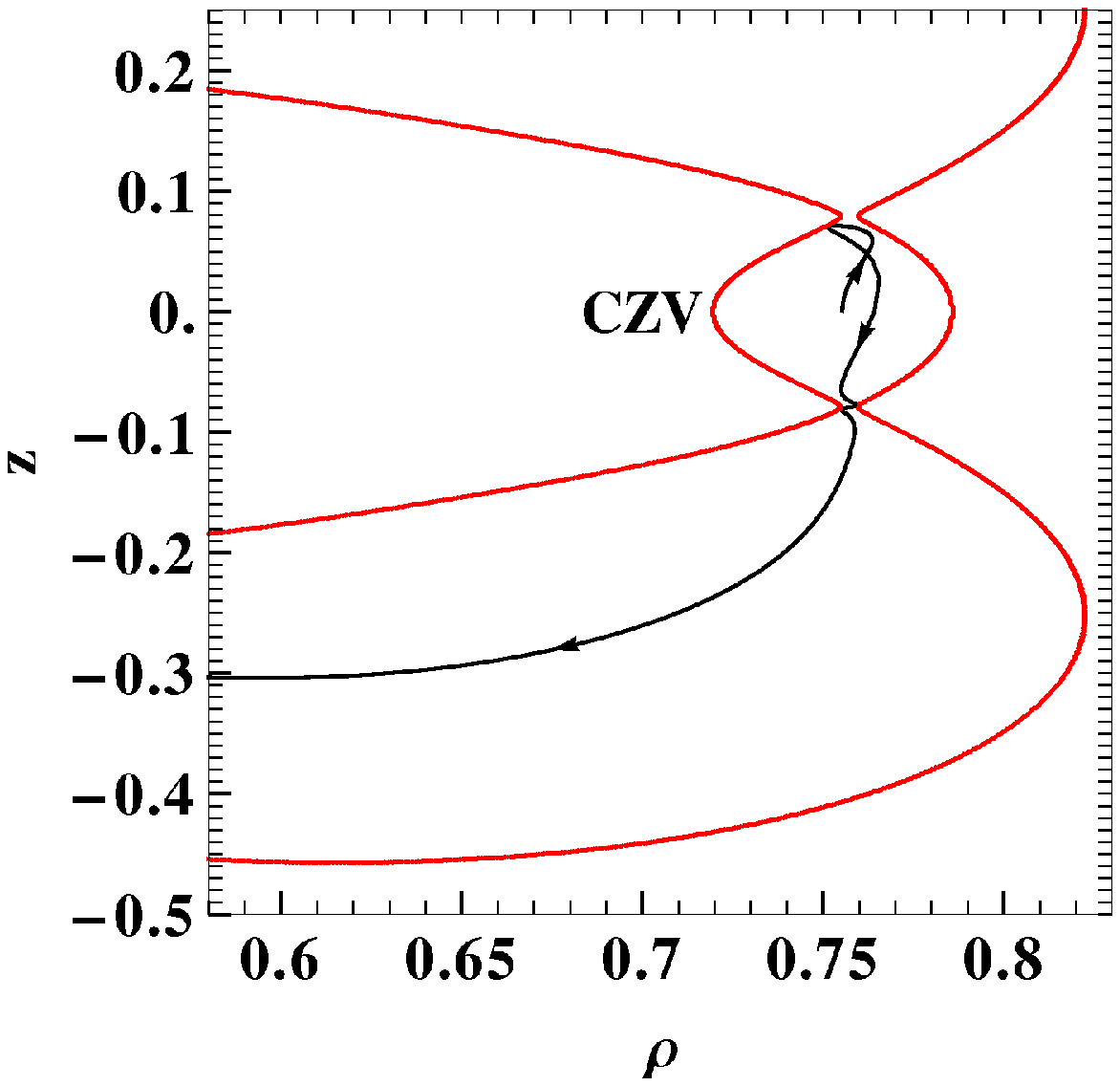}
\caption{\label{FigEsc} The left panel shows characteristics for the energy $E$ of
the families $\mathbf{u}_0^{\prime}$, $\mathbf{u}_0^{\prime\prime}$,
$\mathbf{u}_0^{\prime\prime\prime}$ and of some resonant bifurcations from the
family $\mathbf{u}_0^{\prime\prime}$. The dotted region indicates regions where 
chaos is prominent. The gray region indicates regions with plunging orbits, while
the red region gives the initial conditions (for $z=0$) of orbits that plunge 
directly. The dashed line approximates the position of the first homoclinic section
of the manifolds emanating from the orbit $\mathbf{u}_0^{\prime\prime\prime}$.
The right panel shows a plunging orbit (black curve) and the CZV (red curve).}
\end{minipage}
\end{figure}

A summary of the procedure shown in Fig. \ref{FigSS} is given by the characteristic
curves shown in the left panel of Fig. \ref{FigEsc}. This figure shows different
bifurcations of the orbit $\mathbf{u}_0^{\prime\prime}$ and the creation of a
prominent chaotic region (dotted region) around it as a function of the energy $E$,
while $E$ decreases. When the inner region joins the plunging region, the chaotic
orbits become plunging (gray and red region). This mean that particles of our gas
which move along a chaotic orbit will eventually plunge. Especially, particles
moving along orbits corresponding to the red colored region (left panel of Fig.
\ref{FigEsc}) will plunge directly, i.e., the particles will start from the equator
but they will not cross it again with $\dot{z}>0$ (e.g., in the right panel of Fig.
\ref{FigEsc} the orbit crosses the equator again but with $\dot{z}<0$). The fact
that the red region expands around the prolongation of the characteristic curve of
the $\mathbf{u}_0^{\prime\prime}$ periodic orbit is in agreement with the
aforementioned role of $\mathbf{u}_0^{\prime\prime}$ for creating a path for
chaotic orbits to plunge, because it shows that the plunging paths lie around the
$\mathbf{u}_0^{\prime\prime}$ periodic orbit. Thus, the left panel of Fig.
\ref{FigEsc} shows that the phase space undergoes an abrupt transition. Namely, the
phase space ceases to be dominated by regular orbits and prominent chaotic regions 
appear just before the plunge begins. 

\begin{figure}[htp]
\includegraphics[width=0.4\textwidth]{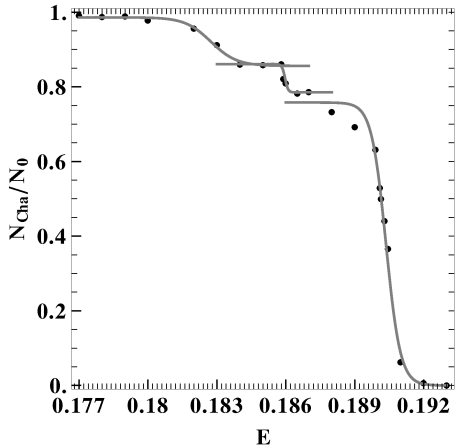}
\hspace{0.1\textwidth}
\begin{minipage}[b]{0.4\textwidth}
\caption{\label{FigChaP} The proportion $N_{cha}/N_0$ of the chaotic-plunging 
orbits with initial conditions along the line $\dot{\rho}=0$ as a function of the
energy $E$.}
\end{minipage}
\end{figure}

This transition is shown better in Fig. \ref{FigChaP}, where we plot the
proportion of chaotic-plunging orbits whose initial condition lie along the line
$\dot{\rho}=0$. The proportion changes abruptly at a narrow interval of decreasing
energy values, and this implies that according to our simplistic model the majority
of the gas particles will plunge along a non-equatorial trajectory, while their
energy lies in this relatively narrow interval. Thus, in a electromagnetic spectrum
coming from the vicinity of the MN central compact body we would expect a
characteristic imprint of such a plunge, because the plunging particles are
creating a non-radial cataclysmic inflow from the accretion disk. Moreover, in Fig.
\ref{FigChaP} we can see two more minor abrupt changes in the proportion
$N_{cha}/N_0$, which means that we may have even some minor inflows as well, which
will have their own characteristic imprints in the spectrum.

Since the particles move outside the equator our simplistic model should hold only
when the enthalpy of the gas is small in comparison to the  gravitational field of
the central object. Otherwise, we have to take into account the gradient of the
pressure of the gas and the geodesic approach has to be further investigated for
its validity. On the other hand, we speculate that adding pressure would be just
like putting a lid over a boiling pot, thus what we expect is that the energy 
interval of the cataclysmic inflow would just become narrower and the abrupt change
in the fraction of chaotic plunging orbits (Fig. \ref{FigChaP}) more steep. Such a
cataclysmic inflow of plunging  matter should have an imprint on the radiation
coming from the surrounding of the central compact object and could, in principle,
be used as a method of detecting whether the black holes are described indeed by
the Kerr metric or not, since such inflows are not expected to exist in the Kerr
case.

\section{Conclusions} \label{sec:concl}

By attempting to give an astrophysical interpretation to a dynamical study
\cite{CHL12}, we suggest a way that a cataclysmic inflow from an initially
non-plunging  region of the accreting matter towards the central compact object may
lead to a method of testing the Kerr hypothesis by checking the electromagnetic
spectra coming from the vicinity of the central compact object.   

\ack G. L-G is supported by the DFG grant SFB/Transregio 7.

\section*{References}
\bibliography{FinalStage}

\end{document}